\documentclass[aps,twocolumn,superscriptaddress,amsfont,graphicx,nofootinbib,preprintnumbers]{revtex4-1}%
\UseRawInputEncoding
\usepackage{amsmath, amsthm, amssymb}
\usepackage{graphicx}
\usepackage{subfigure}
\usepackage{wrapfig}
\usepackage{color}
\usepackage{color,graphicx,epsfig}
\usepackage{ifpdf}
\usepackage{bm}
\usepackage[english]{babel}
\usepackage{braket}
\usepackage{hyperref}
\usepackage{enumerate}
\usepackage{url}
\usepackage{multirow}
\usepackage{tablefootnote} 

\usepackage{slashed}

\usepackage{changes}

\begin{document}
\title{Constraining Bosonic Dark Matter-Baryon Interactions from Neutron Star Collapse}

\author{Chih-Ting Lu}
\email{ctlu@njnu.edu.cn}
\affiliation{Department of Physics and Institute of Theoretical Physics, Nanjing Normal University, Nanjing, 230021, China}

\author{
	Arvind Kumar Mishra 
	${\href{https://orcid.org/0000-0001-8158-6602}{\includegraphics[height=0.15in,width=0.15in]{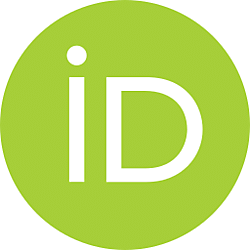}}}$ 
}
\email{arvindm215@gmail.com}
\affiliation{Department of Physics and Institute of Theoretical Physics, Nanjing Normal University, Nanjing, 230021, China}

\author{Lei Wu}
\email{leiwu@njnu.edu.cn}
\affiliation{Department of Physics and Institute of Theoretical Physics, Nanjing Normal University, Nanjing, 230021, China}

\date{\today}

\begin{abstract}
Dark matter (DM) may be captured around a neutron star (NS) through DM-nucleon interactions. We observe that the enhancement of such capturing is particularly significant when the DM velocity and/or momentum transfer depend on the DM-nucleon scattering cross-section. This could potentially lead to the formation of a black hole within the typical lifetime of the NS. As the black hole expands through the accretion of matter from the NS, it ultimately results in the collapse of the host. Utilizing the existing pulsar data J0437-4715 and J2124-3858, we derive the stringent constraints on the DM-nucleon scattering cross-section across a broad range of DM masses. 
\end{abstract}
		

		\maketitle

\section{Introduction}
The presence of dark matter (DM) is firmly established through observational evidence, indicating its gravitational interaction with baryonic matter (see recent review, e.g.~\cite{Bertone:2004pz}). However, despite substantial progress, the nature of the interaction between DM and baryonic matter remains elusive. Ongoing DM searches have yet to reveal any conclusive evidence for its particle nature instead offering constraints on DM-baryon interactions  \cite{Klasen:2015uma,Liu:2017drf,PerezdelosHeros:2020qyt}. DM detection experiments encounter limitations, whether in size or technical constraints \cite{Cebrian:2022brv,MarrodanUndagoitia:2015veg}. Therefore, it becomes imperative to explore extreme cosmic locales that demonstrate sensitivity to the interaction between DM and ordinary matter \cite{Bramante:2023djs}. This exploration promises to shed light on the unexplored realm of DM microphysics.

Neutron stars (NS), characterized by strong gravity and high density, represent exotic environments ideal for investigating the interaction between DM and ordinary baryonic matter~\cite{Zhitnitsky:2023fhs}. Neutron stars situated in DM-dominated regions have the capability to capture surrounding DM particles, consequently modifying their internal structure. The captured DM particles interact with the matter in the NS, undergo thermalization, and can even lead to black hole formation~\cite{Goldman:1989nd, 1990PhLB..238..337G,Kouvaris:2010vv, deLavallaz:2010wp, Bramante:2013nma,Bramante:2013hn, Bell:2013xk, Guver:2012ba, Garani:2018kkd, Kouvaris:2018wnh,Lin:2020zmm,Garani:2021gvc,Singh:2022wvw}. Therefore, the absence of neutron stars in DM-dominated regions serves as a constraint on DM microphysics \cite{Bramante:2014zca}. For a detailed exploration of the impact of DM particles on neutron stars and other compact objects, please refer to Refs.~\cite{Bertone:2007ae,Kouvaris:2010jy,2011PhRvD..84j7301L, Gresham:2018rqo, Deliyergiyev:2019vti,Dasgupta:2020mqg,Dasgupta:2020dik, Kain:2021hpk,Bramante:2022pmn,Bramante:2023djs,Ray:2023auh,Bhattacharya:2023stq,Niu:2024nws} and the references therein. 

In the literature, DM capturing was explored in the presence of constant DM-nucleon scattering cross section~\cite{Zentner:2009is,McDermott:2011jp,Bramante:2014zca}. However, for a more complete treatment, one needs to include the DM velocity and/or momentum transfer depend on the DM-nucleon scattering cross-section. These effects have been explored in the context of the sun~\cite{Guo:2013ypa,Vincent:2013lua,Vincent:2014jia,Vincent:2015gqa,Vincent:2016dcp,Busoni:2017mhe}, white dwarfs~\cite{Steigerwald:2022pjo} and neutron stars~\cite{Bell:2018pkk,Bell:2020lmm,Garani:2020wge,Fujiwara:2022uiq}. However, previous studies are mostly focused on DM capturing and its influence on the thermal conductivity and heating of the star.

In this study, we explore the DM velocity and momentum transfer dependent parametrization for DM-nucleon scattering cross-sections and aim to delve into DM microphysics through the survival analysis of neutron stars. Employing a power-law parametrization for the DM-nucleon scattering cross section in the DM velocity and momentum transfer dependent case, we estimate the DM capture rate and investigate its dependency on DM parameters. Our findings reveal that, in certain scenarios, both DM velocity and momentum transfer dependence lead to an enhanced capturing rate compared to the constant scattering cross-section case.

Following the capture, DM particles interact with the NS materials and gravitate towards the core. We observe that the augmentation of the DM-nucleon scattering cross-section results in a rapid thermalization of the DM particles within the NS. This thermalization suggests the formation of gravitational bound objects, potentially leading to black hole (BH) formation and the destruction of the NS after accreting surrounding material.

Furthermore, by leveraging observational data from neutron stars J0437-4715 and J2124-3858 \cite{Kargaltsev:2003eb,Manchester:2004bp}, we constrain the DM parameters. Our analysis indicates that positive DM velocity dependence enhances DM accretion, imposing stringent constraints on DM parameters compared to the constant scattering cross-section case and momentum-dependent cross-section case. Notably, for some DM mass ranges, the constraints are even stronger than the existing astrophysical and cosmological constraints \cite{Vincent:2015gqa,Vincent:2016dcp,Busoni:2017mhe,Boddy:2018wzy,Ooba:2019erm,Maamari:2020aqz,Buen-Abad:2021mvc}.  

The arrangement of our work is as follows: In section \ref{sec:accretion}, we discuss the capture rate of the DM particles and our parametrization of the DM-Nucleon scattering cross-section. Further, we estimate the number of captured particles and present analytical and numerical results. In Section \ref{therm}, we calculate the thermalization time scale for our models. Then, in Section \ref{BHF}, we explore the black hole formation and discuss the fate of the NS in the light of formed BH accretion and Hawking evaporation. In Section \ref{results}, we present and discuss our main results. Finally, in Section \ref{Sec:conc}, we conclude our work.
\section{\label{sec:accretion} Accretion of DM  on neutron star }
We assume that the neutron star is surrounded by dark matter particles. Here, we focus on bosonic asymmetric dark matter (ADM) so that the annihilation of the DM particles in the current universe can be neglected \cite{Petraki:2013wwa}.  We further assume that DM primarily interacts with the nucleon and neglects any self-interaction within the DM itself. In our analysis, we also neglect the evaporation effect, which is important for $m_{\chi}<2T_{\mathrm{NS}}/v^{2}_{\mathrm{esc}}$  where $T_{\text{NS}}$ and $v_{\mathrm{esc}}$ are the typical temperature and escape velocity of the NS~\cite{McDermott:2011jp}. So, we focus our study on comparably higher DM masses, i.e., $m_{\chi}\geq {\cal{O}}(1)$ keV, where evaporation does not occur. Therefore, using the above assumptions, the number of bosonic ADM particles accumulating in the NS due to elastic DM-neutron interaction is given by
\begin{equation}
\frac{dN_{\chi}}{dt} = C_{\chi\mathrm{N}} ~~,
\end{equation}
where $C_{\chi\mathrm{N}} $ is DM capturing rate due to the DM-neutron interaction. The solution provides us the collected number of the DM particles as $N_{\chi}=C_{\chi\mathrm{N}} ~t$, where $t$ is the accretion time which is typically the lifetime of the NS.

To calculate $N_{\chi}$, we need to provide $C_{\chi\mathrm{N}}$. The calculation of $C_{\chi\mathrm{N}}$ has been extensively discussed in Refs.~\cite{1985ApJ...296..679P,1987ApJ...321..571G,Bell:2020obw,Robles:2022llu,Bell:2023ysh}. In this work, we follow the approach given in Refs. \cite{1985ApJ...296..679P,1987ApJ...321..571G}, in which expression of $C_{\chi\mathrm{N}}$ is given by \footnote{For a more improved estimation of $C_{\chi\mathrm{N}}$, please see Refs.~\cite{Bell:2020obw,Robles:2022llu,Bell:2023ysh} and references therein.}
\begin{equation}
C_{\chi\mathrm{N}}=  \int^{R_{\mathrm{NS}}}_{0} 4\pi  r^{2}  ~\xi \int^{\infty}_{0} \frac{f(u)}{u}~ w \Omega^{-}(w)~du dr~, 
\label{eq:Cnx}
\end{equation}
where $R_{\mathrm{NS}}$ is the radius of NS, $u$ is the DM velocity at very large distance, and $f(u)$ is the DM velocity distribution function. The quantity, $w$ is the DM speed at distance $r$ from the center of the NS, defined by $ w^{2}=u^{2}+v^{2}_{\mathrm{esc}}(r)$, where $v_{\mathrm{esc}}(r)=\sqrt{GM(r)/r}$ is the escape velocity and $M(r)$ is the NS mass enclosed inside the radius $r$. The term $\Omega^{-}(w)$ represents a function related to the scattering kinematics and quantity, $w \Omega^{-}(w)$ is proportional to the probability per unit time of a collision that leads to reduce the speed of DM particles from $w$ to less than $v_{\mathrm{esc}}$. 
In the above equation, $\xi$ represents the nucleon degeneracy effects which depend on the momentum transfer, and given by~\cite{McDermott:2011jp}
\begin{equation}
\xi = \mathrm{min}\left[\frac{q}{p_{\mathrm{F}}},1 \right]~,
\label{eq:xi}
\end{equation}
where $q\simeq \sqrt{2}\left(\frac{m_{\chi}m_{\mathrm{N}}}{m_{\chi}+m_{\mathrm{N}}}\right)v_{\mathrm{esc}}$ is momentum transfer, $m_{\mathrm{N}}$ is the nucleon mass and $p_{\mathrm{F}}=(3\pi^{2}n_{\mathrm{B}})^{\frac{2}{3}}$ is the Fermi energy in which $n_{\mathrm{B}}$ represents the baryon number density. As examples, $v_{\mathrm{esc}}\sim 1.8\times 10^{5}$km$/$sec, $q\geq p_{\mathrm{F}}$ for $m_{\chi}>2$ GeV and $q<p_{\mathrm{F}}$ for $m_{\chi}<2$ GeV.  Therefore, for larger DM masses, we use $\xi =1$. 

Further, we assume that the DM particle velocity distribution function is Maxwellian, given as (in the rest frame of the DM halo) \cite{1985ApJ...296..679P} 
\begin{equation}
 f_{0}(u) = \left(\frac{3}{2} \right)^{\frac{3}{2}} \frac{4}{\sqrt{\pi}} \frac{\rho_{\chi}}{m_{\chi}} \frac{u^{2}}{u^{2}_{0}}e^{-\frac{3}{2}\frac{u^{2}}{u^{2}_{0}}},  
 \end{equation}
 where $\rho_{\chi}$, $m_{\chi}$ represent energy density and mass of the DM. Here $u_{0}=\sqrt{3 T_{\chi}/m_{\chi}}$ is the average velocity and $T_{\chi}$ is the temperature of DM. In the frame of the neutron stars, the DM distribution function becomes 
\begin{equation*}
 f(u) = f_{0}(u) e^{-\frac{3}{2}\frac{u^{2}_{\mathrm{s}}}{u^{2}_{0}}}  \left[\frac{\sinh(3 u u_{\mathrm{s}}/u^{2}_{0})}{(3 u u_{\mathrm{s}}/u^{2}_{0})} \right], 
 \end{equation*}
 where $u_{\mathrm{s}}$ is the velocity of NS with respect to the rest frame of DM halo. In  the dimensionless quantity defined as $x=\left(\frac{3}{2}\frac{u^{2}}{u^{2}_{0}}\right)^{\frac{1}{2}}, ~~ \eta=\left(\frac{3}{2}\frac{u^{2}_{\mathrm{s}}}{u^{2}_{0}}\right)^{\frac{1}{2}}$, and DM distribution function becomes
 \begin{equation}
f(x)=  \frac{4  n_{\chi}}{\sqrt{\pi}} x^{2} e^{-x^{2}-\eta^{2}} \left(\frac{\sinh 2\eta x}{2\eta x} \right),  
\label{distx}
  \end{equation} 
where $n_{\chi}$ represents the DM number density surrounding the NS.

Moreover, to capture DM particles inside of the NS, DM particles need to transfer enough energy to the NS so that after the interaction, their velocity reduces to typically lower than the escape velocity. This condition requires
\begin{equation}
  \frac{u^{2}}{w^{2}}\leq \frac{E_{\mathrm{R}}}{E}\leq \frac{\mu}{\mu^{2}_{+}}~~,
 \end{equation} 
 where $E_{\mathrm{R}}$ is the energy transferred during the DM and nucleon collision, and $E$ is the energy associated with the incoming DM particles. Here $\mu\equiv m_{\chi}/m_{\mathrm{N}}$ and $\mu_{\pm}\equiv (\mu\pm1)/2$ in which $m_{\mathrm{N}}$ is the mass of nucleon. On the other hand, we also assume that a single scattering between the DM and nucleon is sufficient to transfer the energy required for DM capture. This is generally the case for $m_{\chi}<10^{6}$ GeV. However, for higher masses, a single scattering will not be sufficient for enough energy transfer, and multiple scattering is needed for capturing, see Refs. \cite{Steigerwald:2022pjo,Bramante:2017xlb}.

Further, using the above assumptions and considering the interaction cross-section of the DM particles with the nucleons to be constant and isotropic, $\Omega^{-}(w)$ is given by Refs. \cite{1985ApJ...296..679P,1987ApJ...321..571G} 
 \begin{equation}
 \Omega^{-}(w)=n_{\mathrm{N}}\sigma~ w\frac{\mu^{2}_{+}}{\mu}\left( \frac{\mu}{\mu^{2}_{+}} - \frac{u^{2}}{w^{2}}\right) \Theta \left( \frac{\mu}{\mu^{2}_{+}} - \frac{u^{2}}{w^{2}}\right)~,
 \label{eq:omegawg}
 \end{equation}
where $\sigma$ represents the DM-Nucleon scattering cross-section.
Now, we consider a finite-size effect of the nucleon in the DM-Nucleon interaction. This consideration is crucial because, at the NS surface, the DM velocity is quite high, approaching the escape velocity $v_{\mathrm{esc}}$. Consequently, the imparted energy to the nucleon will be large, and DM-Nucleon interaction may no longer behave as a point-like particle. Therefore, DM will interact with the nucleon substructure, i.e., quarks and gluons, and this consideration is relevant for massive DM particles. Therefore, to incorporate this effect, one needs a nucleon form factor in the DM-Nucleon cross-section, i.e., $\sigma=\sigma_{\mathrm{point}}F(q^{2})$, where $\sigma_{\mathrm{point}}$ and $F(q^{2})$ represent the cross-section when the DM and nucleons are point-like and form factor, respectively. In this work, we follow a conservative approach and consider the dipole form factor for DM-nucleon interaction \footnote{In general, the form factor depends on the models of the  DM-nucleon interactions \cite{Cirelli:2013ufw}\cite{Ema:2020ulo} (in context of the NS, see Refs.\cite{Bell:2020obw}\cite{Anzuini:2021lnv}) and references therein. In this work, we have parameterized the DM-Nucleon interaction in the power of relative velocity and momentum transfer, so different models require different form factors. Therefore, to keep our discussion simple and explore the effect of form factor in the capture rate, we adopt a conservative approach and assume a dipole form factor for our every parametrization.} (considered in the next subsection), given by \cite{Ema:2020ulo}
\begin{equation}
F(q^{2})=\frac{\Lambda^{4}}{(q^{2}+\Lambda^{2})^{2}},  
\label{eq:FF}
\end{equation}
where $\Lambda$ is related to the scale below which the nucleon may no longer behave like a point-like object. Considering the typical size of the nucleon is $0.8$ fm \cite{ParticleDataGroup:2022pth}, we assume $\Lambda\simeq 0.25$ GeV throughout our analysis. Our typical estimation suggests that the higher momentum transfer (DM masses) reduces the effective cross-section by two orders of magnitude. 

 \subsection{Parametrization of the DM-baryon elastic scattering}
Now, we discuss our parametrization of the elastic DM-Nucleon interaction. In literature, various types of elastic DM-nucleon scattering have been explored extensively in the context of particle physics \cite{Chang:2009yt,Fan:2010gt,Kumar:2013iva} and cosmology \cite{Vincent:2015gqa,Vincent:2016dcp,Ooba:2019erm,Maamari:2020aqz,Buen-Abad:2021mvc,Boddy:2022tyt}. We mainly focus on the DM-nucleon scattering cross-section, which depends on their relative velocity and momentum transfer during the collision. Here, a model-independent approach is applied, and we parameterize the relative velocity-dependent cross-section (VDCS) as \cite{Vincent:2015gqa}
\begin{equation}
\sigma^{\mathrm{vd}}_{\chi\mathrm{N}}(v_{\mathrm{rel}})=\sigma_{\chi \mathrm{N}}\left(\frac{v_{\mathrm{rel}}}{u_{0}}\right)^{2\alpha}\simeq\sigma_{\chi \mathrm{N}}\left(\frac{w}{u_{0}}\right)^{2\alpha},
\label{eq:vdcs}
\end{equation}
where $v_{\mathrm{rel}}$ is the relative velocity of DM-nucleon scattering and $u_{0}$ is an average DM velocity. Here, $\alpha$ depends on the DM parametrization; however, to keep our analysis general, we consider their representative value $ \alpha=(-1,0,1,2)$ and will not discuss their possible particle physics models. For our calculation purposes, we consider typical value, $u_{0}=220$ km$/$sec. As $w>u_{0}$, the scattering cross-section will increase for positive $\alpha$ and decrease for negative $\alpha$.  

On the other hand, the DM particles can interact with the baryons with the momentum transfer dependent cross-section (MDCS). This kind of scattering cross-section originates in some special types of interaction~\cite{Fan:2010gt,Kumar:2013iva} and can lead to important effects~\cite{Chang:2009yt}. Here, we consider model-independent approach of MDCS parametrized by \cite{Vincent:2015gqa}
\begin{equation}
\sigma^{\mathrm{md}}_{\chi\mathrm{N}}(q)=\sigma_{\chi \mathrm{N}}\left(\frac{q}{q_{0}}\right)^{2\beta},
\label{eq:mdcs} 
\end{equation}
where $q$ is the momentum transfer from DM to the nucleon. To keep our analysis general, we assume some representative values of $\beta$ without discussing their possible particle physics models. In the above, $q_{0}$ is a normalized momentum, and for our purpose, we consider its typical value, $q_{0}=40$ MeV (relevant for the DM direct detection experiment searches \cite{Vincent:2015gqa}).

Finally, we note that the scattering cross-sections can also depend on the combination of the relative velocity and momentum transfer. However, for simplicity, we will ignore this possibility in this study.  

\subsection{Estimation of the captured DM particles}
Now, we calculate the number of collected DM particles, $N_{\chi}$ in the typical lifetime of the NS, $t$, using the velocity and momentum-transfer dependent cross-sections. The analytic forms of $N_{\chi}$ in certain cases are obtained, and for the detail derivations, we refer readers to see appendices \ref{veldep} (for VDCS case) and \ref{momdep} (for MDCS case).
All the analytic expressions are summarized below: 
 
 \textbf{(1) Calculation of $N_{\chi}$ in constant cross-section case:}
For constant cross-section case, the expression for $N_{\chi}$ is obtained as 
 \begin{align}
 & N^{\mathrm{const}}_{\chi}(t)=C^{\mathrm{const}}_{\chi\mathrm{N}}~t =
 \frac{B}{16\eta} P_{0}(A,\mu,\eta)~t
  \nonumber \\
 & = 2.7\times 10^{38} \left(\frac{\rho_{\chi}}{10^{2}\mathrm{GeV~ cm^{-3}}}\right)\left(\frac{\mathrm{GeV}}{m_{\chi}}\right)\left( \frac{\mu^{2}_{-}}{\mu}\right)
 \nonumber \\
 & \times \left(\frac{\sigma_{\chi\mathrm{N}}}{2\times 10^{-45}\mathrm{cm^{2}}}\right) F(q^{2}) P_{0}(A,\mu,\eta)\left(\frac{t}{10^{10}~\mathrm{years}}\right) 
 \label{eq:Nvd0}
 \end{align}
 where $C^{\mathrm{const}}_{\chi\mathrm{N}}$, $B$ and $P_{0}(A,\mu,\eta)$ are given in Eqs. (\ref{eq:Cvd0}),~(\ref{eq:B}) and~(\ref{eq:fvd0}), respectively. In the above equation, $\rho_{\chi}$ is the DM density surrounding the NS, which depends on the position of the star and DM profile adapted. 

\textbf{(2) Calculation of $N_{\chi}$ in VDCS case:}
\begin{itemize}
\item
For $\alpha=1$ (i.e., $\sigma\propto v^{2}_{\mathrm{rel}}$) case, 
\begin{align}
N^{\mathrm{vd}}_{\chi}(\alpha=1,t)=&C^{\mathrm{vd}}_{\chi\mathrm{N}}(\alpha=1)~t=\frac{B}{48 \eta  \mu} P_{1}(A,\mu,\eta) ~t,
\nonumber \\
&=\frac{1}{3 \mu} \frac{P_{1}(A,\mu,\eta)}{P_{0}(A,\mu,\eta)} N^{\mathrm{const}}_{\chi}(t)
 \label{eq:Nvd1}   
\end{align}
 where $C^{\mathrm{vd}}_{\chi\mathrm{N}}(\alpha=1)$ and $P_{1}(A,\mu,\eta)$ are given in Eqs.~(\ref{eq:Cvd1}) and~(\ref{eq:fvd1}), respectively.
 \item
For $\alpha=2$ (i.e., $\sigma\propto v^{4}_{\mathrm{rel}}$) case, 
\begin{align}
N^{\mathrm{vd}}_{\chi}(\alpha=2,t)=&C^{\mathrm{vd}}_{\chi\mathrm{N}}(\alpha=2)~t=\frac{B}{144 \eta  \mu ^2} P_{2}(A,\mu,\eta)~t,
\nonumber \\
&=\frac{1}{9\mu^{2}}\left(\frac{P_{2}(A,\mu,\eta)}{P_{0}(A,\mu,\eta)}\right)N^{\mathrm{const}}_{\chi}(t)
 \label{eq:Nvd1}    
\end{align}
 where $C^{\mathrm{vd}}_{\chi\mathrm{N}}(\alpha=2)$ and $P_{2}(A,\mu,\eta)$ are given in Eqs.(\ref{eq:Cvd2}) and~(\ref{eq:fvd2}), respectively. 
\end{itemize}

\textbf{(3) Calculation of $N_{\chi}$ in MDCS case:}
The number of the collected DM particles in the momentum-transfer dependent case is given by
 \begin{align}
 & N^{\mathrm{md}}_{\chi}(\beta,t)=C^{\mathrm{md}}_{\chi\mathrm{N}}(\beta)~t =
 \left(\frac{q}{q_{0}}\right)^{2\beta}\frac{B}{16\eta} P_{0}(A,\mu,\eta)~t
  \nonumber \\
 & =\left(\frac{q}{q_{0}}\right)^{2\beta} N^{\mathrm{const}}_{\chi}(t)~.
 \label{eq:Nmd1}
 \end{align}
Although the captured DM particles increase with DM-Nucleon scattering cross-section, the cross-section will be maximum at geometric limit characterized by $\sigma_{\mathrm{max}}=\pi R^{2}_{\mathrm{NS}}/N_{\mathrm{B}}$, where $R_{\mathrm{NS}}$ and $N_{B}$ represent the NS radius and the total number of the nucleons in the NS, respectively. Assuming $R_{\mathrm{NS}}=10.6$ km and $N_{B}=1.7\times 10^{57}$, we obtain  $\sigma_{\mathrm{max}}\sim 2\times 10^{-45}$ cm$^{2}$. In the situation when DM-Nucleon scattering cross-section is higher than the geometric limit, the effective value of cross-section saturates, viz $\sigma^{\mathrm{max}}_{\chi\mathrm{N}} =\mathrm{min}\left[\sigma_{\chi\mathrm{N}},\sigma_{\mathrm{max}} \right]$. Therefore, the maximum capture rate is dictated by the saturation limit and estimated as
\begin{equation}
C^{\mathrm{sat}}_{\chi\mathrm{N}}=  \pi R^{2}_{\mathrm{NS}} \int^{\infty}_{0} \frac{f(u)}{u}~ w^{2}~du. 
\label{Cnxs}
\end{equation}
After integration,  we get
\begin{align}
 C^{\mathrm{sat}}_{\chi\mathrm{N}}=  &\frac{\pi}{3}\frac{\rho_{\chi}}{m_{\chi}} R^{2}_{\mathrm{NS}} \bigg[\sqrt{\frac{6}{\pi}}u_{0}e^{-\frac{3}{2}\frac{u^{2}_{\mathrm{s}}}{u^{2}_{0}}}
 \nonumber \\ & +\frac{6G_{\mathrm{N}}M_{\mathrm{NS}}+R_{\mathrm{NS}}(u^{2}_{0}+3u^{2}_{\mathrm{s}})}{R_{\mathrm{NS}}u_{\mathrm{s}}}  \mathrm{erf}\left( \sqrt{\frac{3}{2}}\frac{u_{\mathrm{s}}}{u_{0}}\right)\bigg]   
\end{align}
where $G_{\mathrm{N}}$ is the Newton's gravitational constant, and $M_{\mathrm{NS}}$ is the mass of NS. 
\section{Thermalisation}
\label{therm}
We have seen earlier that the NS accretes the DM particles. The accreted DM particles start interacting with NS materials, get thermalized, and form a self-gravitating system. The energy loss by DM particles via its interaction with the nucleon is  \cite{McDermott:2011jp}
\begin{equation}
\frac{dE}{dt}=-\xi n_{\mathrm{N}}\sigma v \delta E ~,
\end{equation}
where $\xi$, $n_{\mathrm{N}}$, and $v$ represent the neutron degeneracy, nucleon number density, and the relative velocity between the DM and nucleon, respectively. Here $\xi$ is defined by Eq.~(\ref{eq:xi}) and $\delta E = q^{2} /2m_{\mathrm{N}}$ is the energy lost by the DM particles to the nucleon. Therefore, using the above approximation, the thermalization time scale, $t_{\mathrm{th}}$, is obtained as
\begin{equation}
t_{\mathrm{th}}  =  \frac{2m_{\mathrm{N}} E_{\mathrm{th}}}{n_{\mathrm{N}}\xi \sigma v q^{2}}
\label{eq:th}
\end{equation}
where $ E_{\mathrm{th}}$ is the thermal energy. We then replace relative velocity with the thermal velocity obtained as $\frac{1}{2}m_{\chi}v^{2}_{\mathrm{th}}=\frac{3}{2}T=E_{\mathrm{th}},$ so $v_{\mathrm{th}}=\sqrt{\frac{3T}{m_{\chi}}} $, where $T$ is the NS temperature. Furthermore, we point out that if the thermalization time scale is smaller than the lifetime of the NS ($t_{\mathrm{th}}<t_{\mathrm{NS}}$), DM particles will be thermalized inside the NS in its lifetime. 

In case the captured DM thermalizes inside the NS, it forms an isothermal distribution, which is characterized by the thermal radius as
\begin{equation}
r_{\mathrm{th}}=
\left( \frac{9T}{8\pi G_{\mathrm{N}} \rho_{\mathrm{B}} m_{\chi}}\right)^{\frac{1}{2}} ~~.
\end{equation}
where $\rho_{\mathrm{B}}$ is the energy density of the baryons. 

\section{Black hole formation from DM accretion}
\label{BHF}
Now we explore the evolution of DM particles after they are captured inside the NS. If capturing is sufficiently large, a system of DM particles can collapse and form a black hole. Following Ref.~\cite{McDermott:2011jp}, we will briefly discuss the mechanisms for the BH formation and explore the possible fate of neutron stars in the light of formed BH accretion and Hawking evaporation.

\subsection{Black hole formation}
In a system of DM particles, black hole formation occurs when the number of particles becomes larger than the number defined by the Chandrashekhar limit. In this work, we assumed the asymmetric scalar DM \cite{Petraki:2013wwa}. For non-interacting bosonic DM particles, the Chandrashekhar limit is defined as 
\begin{equation}
N^{\mathrm{Boson}}_{\mathrm{Chandra}}\simeq 1.5\times 10^{34} \left( \frac{100 \mathrm{GeV}}{m_{\chi}}\right)^{2} ~~.
\label{eq:Bchand} 
\end{equation}

Here, we will discuss two mechanisms for BH formation: 

\textit{(1) BH formation via Bose-Einstein Condensation (BEC):} We have assumed that DM particles are bosonic; therefore, in the extremely dense region of the NS, DM particles may form a BEC. It happens when the central temperature becomes less than the critical temperature, $T_{c}$. Here $T_{c}$ is defined as 
\begin{equation}
 T_{c}=\frac{2\pi}{m_{\chi}}\left(\frac{n_{\chi}}{\zeta(3/2)} \right)^{\frac{2}{3}} ~.
 \label{eq:crittem}
 \end{equation} 
where $\zeta$ is Riemann-Zeta function. The above equation shows that low mass (high DM number density, $n_{\chi}$) leads to a large critical temperature. In such a situation, BEC formation takes place smoothly. One can define a critical number inside the thermal radius $N^{\mathrm{crit}}_{\chi}=1.06\times 10^{36} \left( T/10^{5}\mathrm{K}\right)^{3}$ by using Eq.~(\ref{eq:crittem}). Therefore, it is clear that whenever the number of captured DM inside the NS is greater than the typical number, $N^{\mathrm{crit}}_{\chi}$, some DM particles will move to the ground state and lead to BEC. The number of particles in the condensed ground state (when $T<T_{c}$) is given by \cite{McDermott:2011jp}
\begin{equation}
N^{0}_{\chi}=N_{\chi} \left[1-\left(\frac{T}{T_{c}} \right)^{\frac{3}{2}}\right]\simeq N_{\chi}-1.06\times 10^{36} \left( T/10^{5}\mathrm{K}\right)^{3}.
 \label{eq:ngrd}
\end{equation}
In the case of BEC, the BH formation will take place when the number of captured DM particles in the condensed ground state is larger than the bosonic Chandrashekhar limit. i.e., $N^{0}_{\chi}>N^{\mathrm{Boson}}_{\mathrm{Chandra}}$. From the above equation, we define a lower limit on $N_{\chi}$, given by
\begin{equation}
N_{\mathrm{BEC}}=N^{\mathrm{Boson}}_{\mathrm{Chandra}}+N^{\mathrm{crit}}_{\chi}~~.
\end{equation}
In this case, the typical mass of the formed BH is given by 
$M_{\mathrm{BH}}\sim m_{\chi}N^{\mathrm{Boson}}_{\mathrm{Chandra}}$.

\textit{(2) BH formation without a BEC (via self-gravitation):} 
We have seen earlier that DM particles may be thermalized inside the NS after the capturing. Further, the system of DM particles starts self-gravitating whenever $\rho_{\chi}$ becomes larger than $\rho_{\mathrm{B}}$ inside the thermal radius. Therefore, the condition for the self-gravitation suggests
\begin{equation}
\rho_{\chi}>\rho_{\mathrm{B}} \quad \mathrm{for} \quad  r \leq r_{\mathrm{th}}.
\end{equation}
Using the above condition, one can define the number of DM particles, $N_{\mathrm{self}}$, required for the self-gravitation inside the NS as
\begin{equation}
N_{\mathrm{self}}\simeq 4.8\times 10^{41} \left( \frac{100 \mathrm{GeV}}{m_{\chi}}\right)^{\frac{5}{2}}  \left( \frac{T}{10^{5} \mathrm{K}}\right)^{\frac{3}{2}}~~.  
\label{eq:Nself}
\end{equation}
 Whenever the accreted DM particles in the NS is greater than this value, i.e., $N_{\chi}\geq N_{\mathrm{Self}}$, DM particles start self-gravitating. In this condition, the BH formation requires $N_{\mathrm{Self}}>N^{\mathrm{Boson}}_{\mathrm{Chandra}}$. According to Eqs.~(\ref{eq:Nself}) and~(\ref{eq:Bchand}), this happens for $m_{\chi}\leq 10^{17} \mathrm{GeV}\left( T/10^{5}\mathrm{K}\right)^{3}$. In this case, the typical mass of the formed BH is given by 
$M_{\mathrm{BH}}\sim m_{\chi}N_{\mathrm{Self}}$.

\begin{figure*}
   \includegraphics[height=2in,width=3.2in]{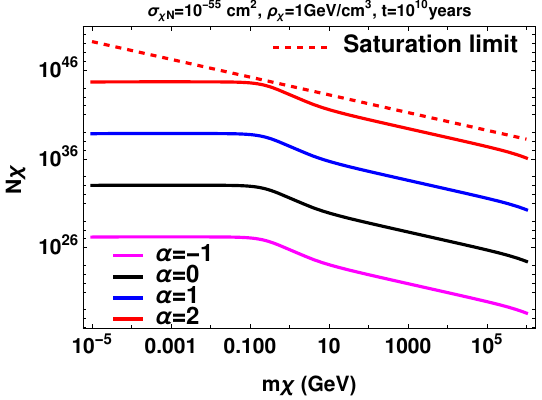}\hfil
 \includegraphics[height=2in,width=3.2in]{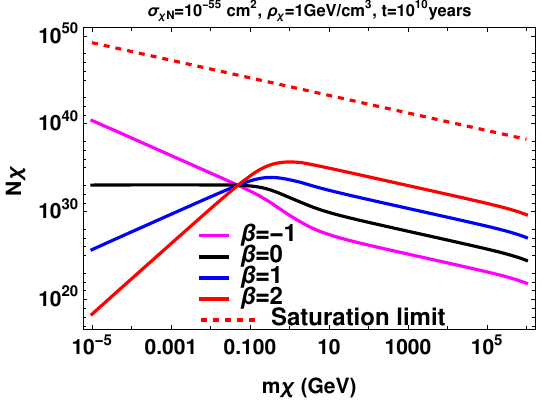}\par
\caption{Effect of the DM velocity dependent (left panel) and DM momentum transfer dependent (right panel) DM-Nucleon scattering cross-section on the DM capture. The red dashed line is the saturation limit of captured DM particles. }
	\label{fig:nvm}%
\end{figure*}
%
\begin{figure*}
   \includegraphics[height=2in,width=3.2in]{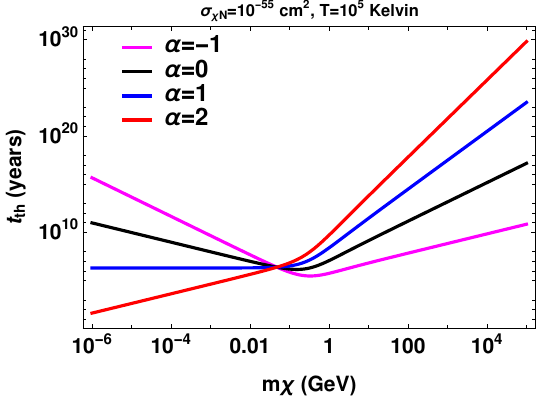}\hfil
 \includegraphics[height=2in,width=3.2in]{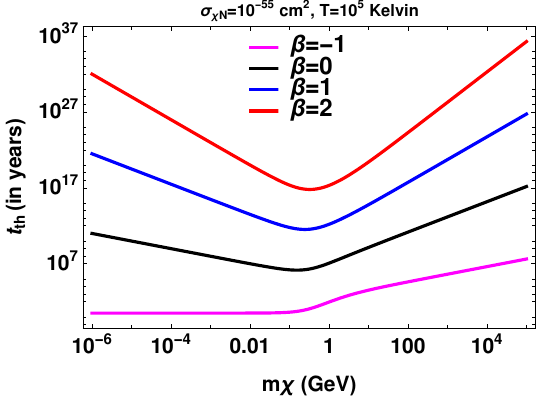}\par
\caption{Effect of the velocity dependent (left panel) and momentum transfer dependent (right panel) DM-Nucleon scattering cross-section on the thermalization scale.  }
	\label{fig:tvm}%
\end{figure*}
\subsection{Stability of the formed black hole}
As soon as the BH forms, two opposite processes start happening simultaneously: first, accretion of the NS material and ambient DM particles surrounding it, which grows BH mass, and second, Hawking evaporation, which reduces BH mass. As the formed BH mass is small (temperature is high), Hawking evaporation plays a dominant role, and if evaporation is large, it can destroy the BH. Therefore, for stability, BH should consume enough material. In light of the above mechanisms, the evolution of the BH with mass $M_{\mathrm{BH}}$ is obtained as 
\begin{align}
 \frac{dM_{\mathrm{BH}}}{dt}\simeq & 4\pi \lambda_{s} \left( \frac{G_{\mathrm{N}}M_{\mathrm{BH}}}{v^{2}_{s}}\right)^{2}\rho_{\mathrm{B}} v_{s}+\left(\frac{dM_{\mathrm{BH}}}{dt}\right)_{\mathrm{DM}} \nonumber \\ 
 &-\frac{1}{1560\pi G^{2}_{\mathrm{N}} M^{2}_{\mathrm{BH}}}~~.
\label{eq:hawking}   
\end{align}
Here, the first term on the right-hand side represents the Bondi-Hoyle accretion rate in which  $\lambda_{s}$, $v_{s}$ correspond to the accretion eigenvalue and sound speed, respectively ~\cite{McDermott:2011jp,Bell:2013xk}. Considering the contribution of the ambient DM accretion is small, from Eq.~(\ref{eq:hawking}), we find that the newly formed BH with the mass $M_{\mathrm{BH}}$ will not be evaporated completely from the Hawking radiation whenever 
\begin{equation}
M_{\mathrm{BH}}>M^{\mathrm{crit}}_{\mathrm{BH}} \simeq \bigg( \frac{v^{3}_{s}}{6240\pi^{2} G^{4}_{\mathrm{N}}\lambda_{s}\rho_{\mathrm{B}}}\bigg)^{\frac{1}{4}}~~.
\label{eq:hawking1}
 \end{equation} 
Assuming the equation of state of neutrons as  $P=K\rho^{\gamma}$, one gets $K=3^{3/2}\pi^{4/3}/(5m^{8/3}_{N})$ and $\lambda_{s}=0.25$ for non-relativistic degenerate neutrons gas case ($\gamma=5/3$) \cite{1983bhwd.book.....S}. From Eq.~(\ref{eq:hawking1}), we find that initial BH with the mass greater than $M^{\mathrm{crit}}_{\mathrm{BH}}=1.2\times 10^{37}$ GeV will grow through the accreting the surrounding materials  
and destroy the NS. Therefore, when $M_{\mathrm{BH}}\sim m_{\chi}N_{\mathrm{Self}}<M^{\mathrm{crit}}_{\mathrm{BH}}$ (when BH form without BEC) and $M_{\mathrm{BH}}\sim m_{\chi}N^{\mathrm{Boson}}_{\mathrm{Chandra}}< M^{\mathrm{crit}}_{\mathrm{BH}}$ (when BH form with BEC), the BH will evaporate.
\section{Results and discussions}
\label{results}
Having equipped ourselves with the basic ingredients, we will now present our main results. Firstly, we will explore the effect of the velocity and momentum transfer dependent cross-section on the capture rate and the thermalization inside the NS. Later, we will constrain the DM parameters by the observations of the neutron stars. For the calculation purposes, we consider $M_{\mathrm{NS}}=1.44 M_{\odot}$, $R_{\mathrm{NS}}=10.6$ km, $N_{\mathrm{B}}=1.7\times 10^{57}$, $u_{0}=220$ km$/$sec, $u_{\mathrm{s}}=230$ km$/$sec, $v_{\mathrm{esc}}=1.8\times 10^{5}$ km$/$sec, and $\rho_{\chi}=0.3\mathrm{GeV~ cm^{-3}}$ otherwise specified.

\subsection{Effect of the VDCS and MDCS on DM capturing}
The number of captured DM particles inside the NS as a function of DM mass is shown in Fig.~\ref{fig:nvm}. The left panel in Fig.~\ref{fig:nvm} corresponds to the velocity-dependent DM-Nucleon scattering cross-section. Here, magenta, black, blue, and red lines correspond to $\alpha=-1, 0, 1, 2$, respectively. The red dashed line corresponds to the saturation limit of the captured DM particles. Here, we find that the number of the collected DM particles, $N_{\chi}$, is constant on the low masses but decreases on the higher masses. The captured DM particles increase for the positive DM velocity-dependent power case but decrease for the negative one. This feature is obvious as $N_{\chi}\propto \sigma$, so the number of the captured DM particles depends on the behavior of the DM-Nucleon scattering cross-section. Further, $N_{\chi}$ estimated for all VDCS cases is always lower than one obtained from the saturation limit.
\begin{figure*}
   \includegraphics[height=2in,width=3.2in]{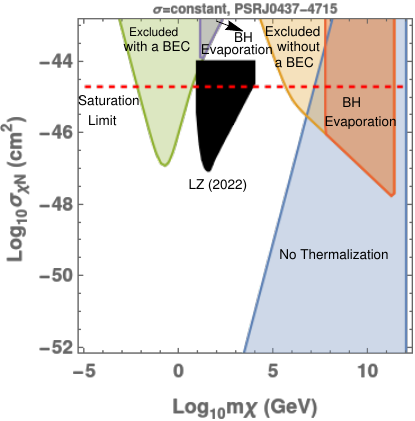}\hfil
 \includegraphics[height=2in,width=3.2in]{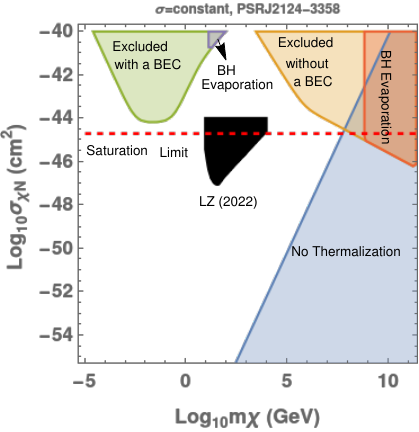}\par
\caption{Constraints on the DM-Nucleon scattering cross-section for constant cross-section case.  The left and right panels are obtained using the pulsar data PSRJ0437-4715 and PSRJ2124-3858, respectively. Here, olive, brown, and sky-blue regions are excluded from the BH formation with a BEC, without a BEC, and non-thermalization of the captured DM particles. The red dashed line corresponds to the saturation limit of the thermalization time scale.  Further, the black region corresponds to the constraint on the DM-Nucleon scattering cross-section obtained from the LUX-ZEPLIN (LZ) experiment~\cite{LZ:2022lsv} for the comparison, and purple and orange regions represent the BH evaporation. The red dashed line is the saturation limit of the DM-Nucleon scattering cross-section. }
\label{fig:ind}%
\end{figure*}

The right panel in Fig.~\ref{fig:nvm} shows the captured DM particles as a function of DM mass in the MDCS case. Magenta, black, blue, and red lines correspond to $\beta=-1, 0, 1, 2$, respectively. As mentioned earlier, the red dashed line corresponds to the saturation limit of the captured DM particles. Here, we find that collected DM particles decrease for high DM masses for all powers.  However,  on low DM masses, unlike the VDCS case (where $N_{\chi}$ is constant for all powers), $N_{\chi}$ decreases sharply for positive powers,i.e., $\beta=1, 2$ and increases for $\beta=-1$ case. The striking feature of the MDCS case is that for low-mass DM particles, the captured rate is smallest for $\beta=2$ and largest for $\beta=-1$. 
\subsection{DM-thermalization}
Fig. \ref{fig:tvm} shows the thermalization time scale as a function of DM mass. The left and right panels correspond to the velocity dependent and momentum transfer dependent DM-Nucleon scattering cross-section, respectively. Here we see that for the velocity dependent cases, $t_{\mathrm{th}}$ increases with the DM mass. In comparison with the constant cross-section, the time scale is smaller for low DM masses, whereas it is larger for high DM masses. On the contrary, for the $\alpha=-1$ case, the thermalization time scale is largest and smallest on the low and high DM mass among all VDCS cases, respectively. In the momentum transfer dependent case, the thermalization scale is largest for $\beta=2$ and smallest for the $\beta=-1$ case. This is true for all DM mass ranges. 
%
\begin{figure*}
   \includegraphics[height=2in,width=3.2in]{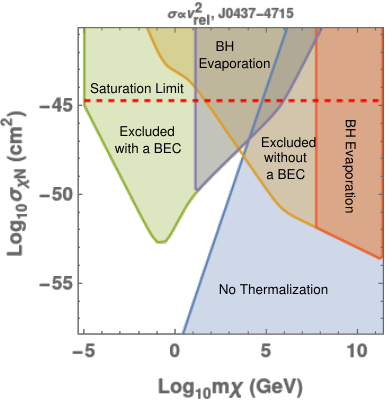}\hfil
 \includegraphics[height=2in,width=3.2in]{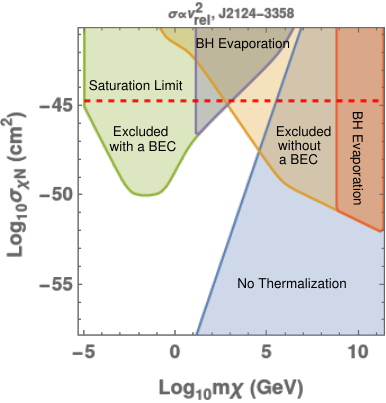}\par
\caption{Contraints on the DM-Nucleon scattering cross-section in VDCS with $\alpha=1$. The left and right panels are obtained from the pulsar data PSRJ0437-4715 and PSRJ2124-3858, respectively. Here, the colored regions carry the same meanings as Fig.~\ref{fig:ind}.}
	\label{fig:vd1}%
\end{figure*}
%
\subsection{Constraints on the DM-Nucleon interaction in the velocity dependent cross-section }
\label{subsec:vdconst}
Now, we will constrain the DM parameters from the survival of the NS (i.e., the gravitational collapse of the DM particles will not destroy the host NS). However, before that, we will briefly summarize the conditions that restrict the BH formation inside the neutron stars following Ref.~\cite{McDermott:2011jp}. When the BEC formation does not occur, the BH formation is preferable on the large DM masses. The survival of the NS requires, $N_{\chi}<N^{\mathrm{Bosons}}_{\mathrm{Chandra}}$. In terms of $N_{\mathrm{Self}}$ this condition suggests 
\begin{equation}
N_{\chi}<N_{\mathrm{Self}}, \quad \mathrm{valid~ for} \quad m_{\chi}\leq 10^{17}\left( T/10^{5}\mathrm{K}\right)^{3}
\end{equation}
as $N_{\mathrm{Self}}>N^{\mathrm{Bosons}}_{\mathrm{Chandra}}$ for $m_{\chi}\leq 10^{17}\left( T/10^{5}\mathrm{K}\right)^{3}$. In cases where the BEC formation occurs, the BH formation is preferable on the small DM masses. Here the servival of the neutron stars requires, $N^{0}_{\chi}<N^{\mathrm{Bosons}}_{\mathrm{Chandra}}$. Using this, in terms of collected DM numbers, the condition for the survival of the NS becomes
\begin{equation}
N_{\chi}<N_{\mathrm{BEC}}~.
\end{equation}
We then consider the pulsar data and constrain the DM parameters. Here, two cold and old neutron stars with known parameters are involved. First, PSRJ0437-4715 at distance $139\pm 3$ pc, surface temperature, $T_{e}=1.2\times 10^{5}$K 
\cite{Kargaltsev:2003eb}, $t_{\mathrm{age}}\sim 6.69\times 10^{9} $ years 
\cite{Manchester:2004bp}. Second, PSRJ2124-3858 at distance $270$ pc, surface temperature, $T_{e}<4.6\times 10^{5}$K 
\cite{Kargaltsev:2003eb}, $t_{\mathrm{age}}\sim 7.81\times10^{9} $ years 
\cite{Manchester:2004bp}. The central temperature can be obtained from the analytic formula given in Ref. \cite{1982ApJ...259L..19G} using the surface data. This provides us $T_{c}=2.1\times 10^{6}$K and $T_{c}=2.5\times 10^{7}$K for the neutron stars PSRJ0437-4715 and PSRJ2124-3858, respectively ~\cite{McDermott:2011jp}. We will constrain the DM parameters from the above two NS data.

\begin{figure*}[]
\includegraphics[height=2in,width=3.2in]{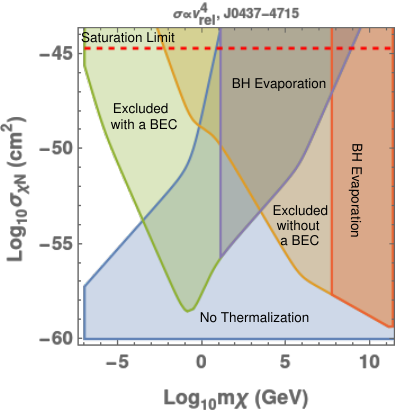}\hfil
\includegraphics[height=2in,width=3.2in]{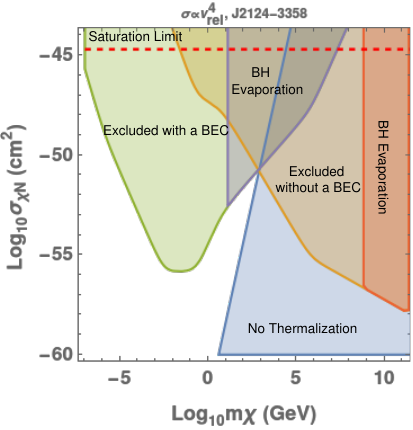}\par
\caption{Constraints on the DM-Nucleon scattering cross-section for VDCS with $\alpha=2$ case. The left and right panels are obtained from the pulsar data PSRJ0437-4715 and PSRJ2124-3858, respectively. Here, the colored regions carry the same meanings as Fig.~\ref{fig:ind}.}
	\label{fig:vd2}%
\end{figure*}
Fig.~\ref{fig:ind} shows the constraint on the bosonic DM-Nucleon scattering cross-section for the constant cross-section case by using the pulsar data, J0437-4715 (left panel) and J2124-3858 (right panel). The olive and brown regions are excluded from the BH formation with a BEC and BH formation without a BEC. Further, the purple and orange colors represent the regions where the BH evaporation dominates over the accretion rate, and BH loses its mass with time. Therefore, in these regions, the constraints are relaxed. In the sky-blue region, captured DM particles do not thermalize with the neutrons and cannot distribute themselves within the thermal radius. Hence, in this region, bounds will not be applicable. In addition, the black region is the constraint obtained from the LUX-ZEPLIN (LZ) experiment~\cite{LZ:2022lsv} for comparison \footnote{It is important to emphasize that the LZ (2022) constraint~\cite{LZ:2022lsv} is specifically shown for the constant cross-section case. However, as there are no explicit constraints (using the LZ(2022) data) available for velocity dependent and momentum transfer dependent DM-Nucleon scattering cross-section cases in the literature, we will not compare our constraints (obtained from VDCS and MDCS scenarios) with the LZ(2022)  data.}, and the red dashed line corresponds to the saturation limit of the DM-Neucleon scattering cross-section.  

From Fig.~\ref{fig:ind}, we find that pulsar data J0437-4715, constrains the low DM mass range, $5\times 10^{-3} \mathrm{GeV}~\leq m_{\chi} \leq 10^{2}$ GeV from BH formation with a BEC. The constraint on higher DM masses $m_{\chi} \geq 5$ GeV are relaxed from the BH evaporation (purple region), and saturation limit on cross-section. Hence, J0437-4715 constrains, $5\times 10^{-3} \mathrm{GeV}~\leq m_{\chi} \leq 5$ GeV from a BH formation via BEC. Further, J0437-4715 constrains the higher DM masses, $m_{\chi} \geq 3\times 10^{5}$ GeV from the BH formation without a BEC. However, due to the BH evaporation, constraints are relaxed for $m_{\chi} \geq 1\times 10^{7}$ GeV (orange region). Therefore, using the evaporation and thermalization constraints, J0437-4715 limit, $3\times 10^{5}\mathrm{GeV}~ \leq m_{\chi} \leq  10^{7}$ GeV from the BH formation without a BEC. On the other hand, pulsar J2124-3858 data indicates that the BH formation with a BEC requires a DM-nucleon scattering cross-section larger than saturation limits; therefore, this data rules out the possibility of the BH formation via BEC. Further, the pulsar data constrains $m_{\chi} \geq 1.5\times 10^{8}$ GeV; however, these constraints are relaxed from the non-thermalization and the BH evaporation.  Therefore, J2124-3858 data does not constrain the DM physics in this model. In light of the above, we point out that the constraints on $\sigma_{\chi\mathrm{N}}$ are weaker in comparison with the constant cross-section case where the form factor has not been taken into account (see Fig. 2 in Ref.~\cite{McDermott:2011jp}).

\begin{figure*}[]
   \includegraphics[height=2in,width=3.2in]{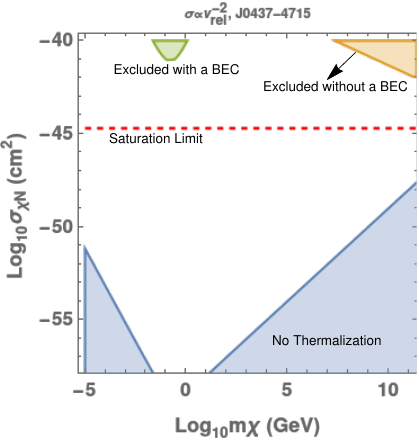}\hfil
 \includegraphics[height=2in,width=3.2in]{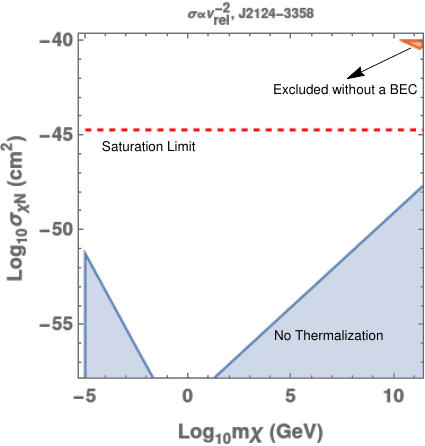}\par
\caption{Constraints on the DM-Nucleon scattering cross-section for VDCS with $\alpha=-1$ case. The left and right panels are obtained from the pulsar data PSRJ0437-4715 and PSRJ2124-3858, respectively. Here, the colored regions carry the same meanings as Fig.~\ref{fig:ind}.}
	\label{fig:vdm1}%
\end{figure*}
\begin{figure*}
   \includegraphics[height=2in,width=3.2in]{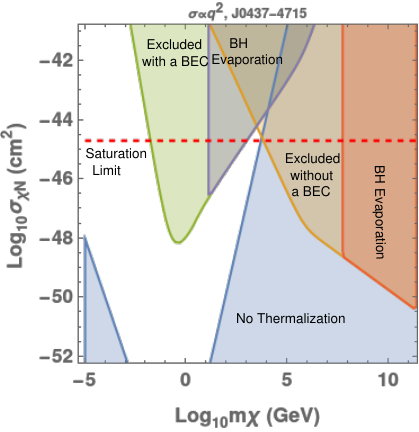}\hfil
 \includegraphics[height=2in,width=3.2in]{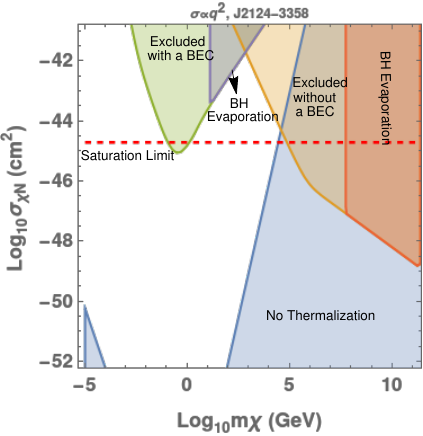}\par
\caption{Constraints on the bosonic DM-Nucleon scattering cross-section for MDCS with $\beta=1$ case. The left and right panels are obtained from the pulsar data PSRJ0437-4715 and PSRJ2124-3858, respectively. Here, the colored regions carry the same meanings as Fig.~\ref{fig:ind}.}
	\label{fig:md1}%
\end{figure*}
Fig.~\ref{fig:vd1} shows the constraint on the bosonic DM-Nucleon scattering cross-section for the VDCS with $\alpha=1$ by using the pulsar data, J0437-4715 (left panel) and J2124-3858 (right panel). The red-dotted line and colored regions carry the same meanings as Fig.~\ref{fig:vd1}. Here, we see that J0437-4715 constrains, $ 10^{-5}\mathrm{GeV}~ \leq m_{\chi} \leq 10^{6}$, from the BH formation with a BEC. However, due to the BH evaporation, constraints are relaxed for $m_{\chi} \geq 17$ GeV (purple region), so J0437-4715 constrains $10^{-5}\mathrm{GeV}~\leq m_{\chi} < 17$ GeV from BH formation with a BEC. Further, from the BH formation without a BEC case, J0437-4715 data constrains $m_{\chi} \geq 38$ GeV; nevertheless, due to the BH evaporation and thermalization constraint, parameter space is limited, $38\mathrm{GeV}~ \leq m_{\chi} \leq 5.7\times 10^{3}$ GeV. Further, J2124-3858 data constrains the DM parameters, $10^{-5}\mathrm{GeV}~\leq m_{\chi} < 16$ GeV and $4.3\times 10^{2}\mathrm{GeV}~ \leq m_{\chi} \leq 3.5\times 10^{4}$ GeV from BH formation with a BEC and without a BEC, respectively. In this case, due to the positive DM velocity dependence ($\sigma\propto v^{2}_{\mathrm{rel}}$), the cross-section and collected DM number density increase sharply, and the possibility of the BH formation becomes high. Therefore, in this scenario, the DM parameter constraints are stronger than the constant cross-section case (see Fig. 2 in Ref.~\cite{McDermott:2011jp}).

In Fig.~\ref{fig:vd2}, we show the constraint on the bosonic DM-Nucleon scattering cross-section for VDCS with $\alpha=2$ from the pulsar data J0437-4715 (left panel) and J2124-3858 (right panel). The red-dotted line and colored regions carry the same meanings as earlier. Here we find that J0437-4715 constrains $ 10^{-7} ~\mathrm{GeV} \leq m_{\chi} \leq 12.7$ GeV from the BH formation with a BEC and $5\times 10^{-3} ~\mathrm{GeV} \leq m_{\chi} \leq 6.6\times 10^{-2}$ GeV from the BH formation without a BEC. Further, pulsar data J2124-3858 contrains $10^{-7} ~\mathrm{GeV} \leq m_{\chi} \leq 12$ GeV from the BH formation with a BEC and $5\times 10^{-2} \mathrm{GeV} \leq m_{\chi} \leq 7.5\times 10^{2}$ GeV from the BH formation without BEC. In this scenario, cross-section again depends on the positive power of DM velocity but more strongly than in the previous case; therefore, the collected DM density and BH formation probability will be even higher. Therefore, the DM mass constraint for this scenario is stronger than that of the constant cross-section and quadratic DM velocity dependent one.

  \begin{figure*}
\includegraphics[height=2in,width=3.2in]{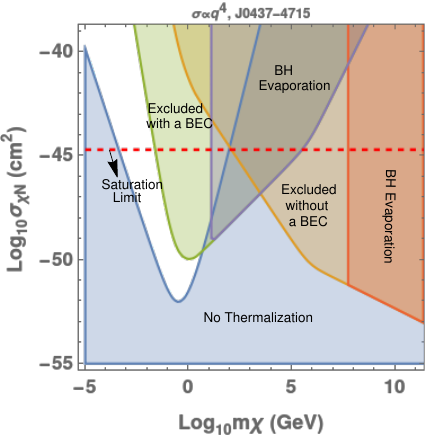}\hfil
\includegraphics[height=2in,width=3.2in]{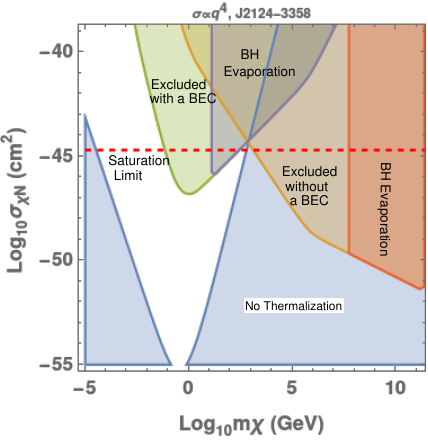}\par
\caption{Constraints on the bosonic DM-Nucleon cross-section for MDCS with $\beta=2$ case. The left and right panels are obtained using the pulsar data PSRJ0437-4715 and PSRJ2124-3858, respectively.  Here, the colored regions carry the same meanings as Fig.~\ref{fig:ind}.}
	\label{fig:md2}%
\end{figure*}

 %
\begin{figure*}
\includegraphics[height=2in,width=3.2in]{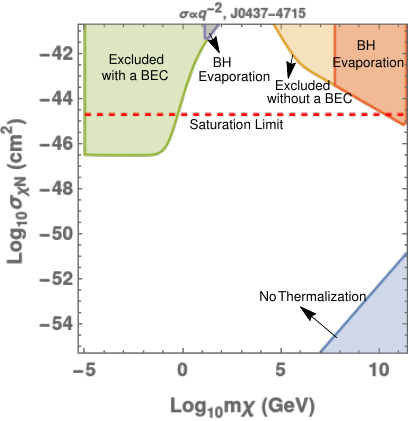}\hfil
\includegraphics[height=2in,width=3.2in]{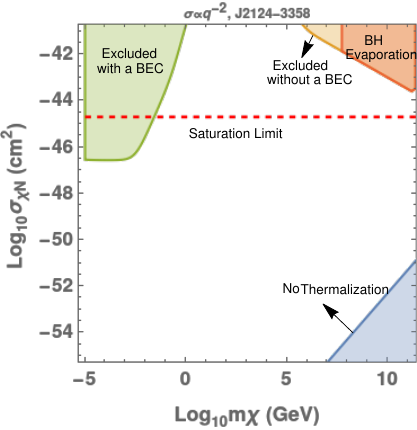}\par
\caption{Constraints on the bosonic DM-Nucleon cross-section for MDCS with $\beta=-1$. The left and right panels are obtained from the pulsar data PSRJ0437-4715 and PSRJ2124-3858, respectively. Here, the colored regions carry the same meanings as Fig.~\ref{fig:ind}.}
	\label{fig:mdm1}%
\end{figure*}
%
Finally, Fig.~\ref{fig:vdm1} shows the constraint on the bosonic DM-Nucleon scattering cross-section for VDCS with $\alpha=-1$ from the pulsar data discussed above. Again, the color of the plotted regions carries the same meanings as earlier. Interestingly, accreted DM particles are insufficient for both neutron stars to form black holes.
This is because of the negative DM velocity dependence, the cross-section, and hence, the collected DM number decreases (compared to the constant cross-section). It reduces the possibility of BH formation, which requires a cross-section larger than the saturation limit. Therefore, this scenario fails to constrain DM parameters from the BH formation.

Furthermore, we emphasize that for the VDCS case, the accreted DM particles are increased for the positive power, which enhances the probability of BH formation in the host NS. Our analysis shows that the pulsar data provides a strong constraint on the DM parameters for low mass DM particles, i.e., $ 10^{-7} ~\mathrm{GeV} \leq m_{\chi} \leq 12.7$ GeV for model $\sigma\propto v^{4}_{\mathrm{rel}}$.  In addition, the scenario $\sigma\propto v^{2}_{\mathrm{rel}}$ constrains the intermediate DM masses, $4.3\times 10^{2}\mathrm{GeV}~ \leq m_{\chi} \leq 3.5\times 10^{4}$ GeV and the constant cross-section constrains the high mass DM particles, $3\times 10^{5}\mathrm{GeV}~ \leq m_{\chi} \leq  10^{7}$ GeV.  We point out that these constraints are more stringent than the existing astrophysical and cosmological constraints \cite{Vincent:2015gqa,Vincent:2016dcp,Busoni:2017mhe,Boddy:2018wzy,Ooba:2019erm,Maamari:2020aqz,Buen-Abad:2021mvc}.  
 \subsection{Constraints on the DM-Nucleon interaction in the momentum dependent cross-section}
 \label{subsec:mdconst}
In this subsection, we constrain the bosonic DM-Nucleon scattering cross-section in the MDCS from the survival of the NS. Fig.~\ref{fig:md1} shows the constraint on the DM-Nucleon scattering cross-section for MDCS with $\beta=1$ from the pulsar data J0437-4715 (left panel) and J2124-3858 (right panel). The color of the plotted regions carries the same meanings as Fig.~\ref{fig:vd1}. We see that J0437-4715 constrains $0.01 \mathrm{GeV} \leq m_{\chi} \leq 13$ GeV from the BH formation via BEC but does constrain the DM parameters from the BH formation without a BEC.
Further, J2124-3858 constrains a small DM mass range, i.e., $0.1 \mathrm{GeV} \leq m_{\chi} \leq1.2$ GeV from BH formation with a BEC but it fails to constrain the DM parameters from the BH formation without a BEC.

 Fig.~\ref{fig:md2} shows the constraints on the DM-Nucleon scattering cross-section for MDCS with $\beta=2$ from the pulsar data discussed earlier. We see that pulsar data J0437-4715 and J2124-3858 constrain $0.06 \mathrm{GeV} \leq m_{\chi} \leq 13$ GeV and $0.02 \mathrm{GeV} \leq m_{\chi} \leq 12.6$ GeV from the BH formation via BEC. However, for both pulsar data, DM parameters are unconstrained from the BH formation without a BEC.

 Finally, Fig.~\ref{fig:mdm1} shows the constraints on the DM-Nucleon scattering cross-section for MDCS with $\beta=-1$ using the pulsar data. Here, we see that BH formation occurs specifically at low DM masses, not at high DM masses. It is because, for this scenario, the accreted DM particles increase for low DM masses and decrease quickly for large DM mass (see right panel of Fig. \ref{fig:nvm}). The pulsar data J0437-4715 and J2124-3858 constrains $ m_{\chi} \leq 0.6$ GeV and $ m_{\chi} \leq 0.03$ GeV from the BH formation via BEC. Further, similar to the previous case, BH does not form without a BEC, so the DM parameters are unconstrained. 
 
Therefore, we emphasize that in MDCS case, pulsar data constrains the low mass DM particles,$ m_{\chi} \leq 0.6$ GeV for $\sigma\propto q^{-2}$ and slightly higher mass DM particles, $0.02 \mathrm{GeV} \leq m_{\chi} \leq 13$ GeV for the positive power of momentum transfer dependence, i.e., $\sigma\propto q^{2}$ and $\sigma\propto q^{4}$.
\section{Conclusion}
\label{Sec:conc}

Due to strong gravity and high density, neutron stars (NS) are a lucrative place to explore dark matter (DM) microphysics.  Neutron stars situated in the DM-rich environment can capture the DM particles. The captured DM particles can be thermalized, form a gravitationally bound core, and finally form a black hole inside the NS.  After accreting the surrounding material, the black hole can destroy the NS. In this work, we constrain the DM microphysics using the data from the survival of the NS.

Assuming the DM velocity and momentum transfer dependent parametrization of the DM-nucleon scattering cross section, we estimated the capture rate of the DM particles by the NS. In a few cases, both the DM velocity and momentum transfer dependence on DM-nucleon scattering cross-section led to enhancement in the capturing rate over the constant scattering cross-section case. We report that the enhancement of DM-nucleon scattering cross-section leads to quick thermalization of the DM particles inside the NS.

Further, using observational data of the neutron stars J0437-4715 and PSRJ2124-3858,  we have constrained the DM parameters. In our analysis, we reported that the positive DM velocity dependence enhances the DM accretion rate and, therefore, strongly constrains the DM parameter comparison to the constant scattering cross-section case. We found that the velocity dependent cross-section imposes the most stringent constraints on DM parameters. Specifically, $\sigma\propto v^{4}_{\mathrm{rel}}$ constrains low mass DM particles, ($ 10^{-7} ~\mathrm{GeV} \leq m_{\chi} \leq 12.7$ GeV), $\sigma\propto v^{2}_{\mathrm{rel}}$ constrains the intermediate DM masses, ($4.3\times 10^{2}\mathrm{GeV}~ \leq m_{\chi} \leq 3.5\times 10^{4}$ GeV ), and the constant cross-section constrains the high mass DM particles, ($3\times 10^{5}\mathrm{GeV}~ \leq m_{\chi} \leq  10^{7}$ GeV).  Notably, these constraints are stronger than existing astrophysical and cosmological constraints. Furthermore, we obtained that the momentum transfer dependent cross-section places restrictions on low-mass DM particles ($m_{\chi} \leq 0.6$ GeV) for $\sigma\propto q^{-2}$, and on comparatively higher mass DM particles ($0.02$ GeV $\leq m_{\chi} \leq 13$ GeV) for positive powers of momentum transfer dependence, i.e., $\sigma\propto q^{2}$ and $\sigma\propto q^{4}$.

Furthermore, we also pointed out that in this work, we assumed certain assumptions, such as the system is Newtonian, non-relativistic, baryon and DM densities are constant inside the NS and escape velocity is constant.  Therefore, when all effects are taken into account, the constraint will be modified. In the future, we would like to explore these aspects and apply our refined formalism to investigate the properties of DM from other compact objects.

\section{ACKNOWLEDGEMENT}
This work is supported by the National Natural Science Foundation of China (NNSFC) under grants No. 12335005, No. 12275134, and No. 12147228. AKM would like to thank Liangliang Su for the useful discussions. 

\appendix
\section{Derivation of accreted DM particles via VDCS}
\label{veldep}
From Eq.(\ref{eq:Cnx}), it is clear that the integration involves the velocity, $u$, and $r$ integration. To simplify this, we assume that the DM and nucleon densities are constant inside the NS, i.e., $n_{\mathrm{N}}(r)=n_{\mathrm{N}}$, and $n_{\chi}(r)=n_{\chi}$. Further as the variation of escape velocity also depends on NS density, so for simplicity we neglect the variation of the escape velocity with $r$ and consider its value at the surface on the NS, i.e., $v_{\mathrm{esc}}(r)\sim v_{\mathrm{esc}}(R_{\mathrm{NS}})=v_{\mathrm{esc}}$.  In this case the cross-section depends on the DM velocity, and to simplify this, we rewrite the $\sigma^{\mathrm{vd}}_{\chi\mathrm{N}}$ in terms of the dimensionless variables, $x$ and $A=\left(\frac{3}{2}\frac{u^{2}}{u^{2}_{0}} \frac{\mu}{\mu^{2}_{-}} \right)^{\frac{1}{2}}$. After the simplification, we obtain $w(x)=\left[\frac{2u^{2}_{0}}{3}\left(x^{2}+\frac{A^{2}\mu^{2}_{-}}{\mu}\right)\right]^{\frac{1}{2}}$,  and 
\begin{equation}
\sigma^{\mathrm{vd}}_{\chi\mathrm{N}}\simeq\sigma_{\chi\mathrm{N}}\left(\frac{w(x)}{u_{0}}\right)^{2\alpha}=\sigma_{\chi\mathrm{N}}\left[\frac{2}{3} \left(x^{2}+\frac{A^{2}\mu^{2}_{-}}{\mu} \right)\right]^{\alpha}.
\end{equation}
 Further, in this case, $\Omega^{-}(w)$ is given by  
\begin{equation}
 \Omega^{-}(w)=n_{\mathrm{N}}\sigma^{\mathrm{vd}}_{\chi\mathrm{N}}F(q^{2})~w~\frac{\mu^{2}_{+}}{\mu}\left( \frac{\mu}{\mu^{2}_{+}} - \frac{u^{2}}{w^{2}}\right) \Theta \left( \frac{\mu}{\mu^{2}_{+}} - \frac{u^{2}}{w^{2}}\right).
  \label{eq:omegawvd}
 \end{equation}
In terms of the dimensionless variables, the above equation is simplified as 
 \begin{equation}
 \Omega^{-}(w)= n_{N}~ \sigma^{\mathrm{vd}}_{\chi\mathrm{N}} F(q^{2})~\frac{1}{w}~\frac{v^{2}}{A^{2}}~\left(A^{2} -x^{2} \right)\Theta(A-x).
 \label{eq:omegavd1}
 \end{equation}

Having been equipped with the basic equations, we can now apply Eq.(\ref{eq:Cnx}), Eq.(\ref{distx}), and Eq.(\ref{eq:omegavd1}) to obtain the capture rate of the DM particle.  We emphasize that even after this simplification, it is difficult to get the general form of the $C_{\chi\mathrm{N}}$ for arbitrary velocity-dependent cross-section, i.e., $C^{\mathrm{vd}}_{\chi\mathrm{N}}(\alpha)$ but for some special cases, viz. $ \alpha=(0,1,2)$, we can get the analytic results.   
\subsection{For constant cross-section case ($\sigma=\sigma_{\chi\mathrm{N}}$)}
To obtain the $C_{\chi\mathrm{N}}$ for the constant DM-Nucleon scattering cross-section, we consider $\alpha=0$. The analytic expression of $C_{\chi\mathrm{N}}$ is given by 
\begin{equation}
C^{\mathrm{const}}_{\chi\mathrm{N}}=\frac{B}{16\eta} f_{0}(A,\mu,\eta), 
\label{eq:Cvd0}
\end{equation}
where 
\begin{equation}
\quad B= \sqrt{\frac{32}{3}}\frac{\mu^{2}_{-}}{\mu}F(q^{2}) N_{\mathrm{B}}n_{\chi}u_{0}\sigma_{\chi \mathrm{N}}~~,
\label{eq:B}
\end{equation}
 \begin{align}
& \mathrm{and} \quad P_{0}(A,\mu,\eta) = \nonumber \\
&\sqrt{\pi } \left(2 A^2-2 \eta ^2-1\right) \text{erf}(A-\eta )
 \nonumber \\ &
 +e^{-(A+\eta )^2}\bigg[2 A \left(e^{4 A \eta }-1\right)+ 2 \eta  \left(e^{4 A \eta }-2 e^{A (A+2 \eta )}+1\right) \nonumber \\ & +  \sqrt{\pi }e^{(A+\eta )^2} \left(2 A^2-2 \eta ^2-1\right) (\text{erfc}(A+\eta )-2 \text{erfc}(\eta )+1)\bigg], 
 \label{eq:fvd0}
 \end{align}
where $\mathrm{erf(z)}$ and $\mathrm{erfc(z)}$ represent the error function, and complimentary error function, defined via $\mathrm{erf(z)}=\frac{2}{\sqrt{2}} \int^{z}_{0}e^{-y^{2}}dy$, and $\mathrm{erfc(z)}=1-\mathrm{erf(z)}$, respectively.
In the above equation, $n_{\chi}$ is the number density of the DM surrounding the NS, and $N_{\mathrm{B}}$ is the total number of baryons inside the NS. We checked that in the rest frame of DM halo, i.e., $\eta \rightarrow 0$, and in an assumption of the point-like structure of nucleon (neglecting the form factor), our expression of $C^{\mathrm{const}}_{\chi\mathrm{N}}$ matches with the Ref.~\cite{1987ApJ...321..571G}. 

\subsection{For $\alpha=1$ (i.e., $\sigma\propto v^{2}_{\mathrm{rel}}$) case}
 For $\alpha=1$, an analytic expression of $C_{\chi\mathrm{N}}$ is given by 
\begin{equation}
C^{\mathrm{vd}}_{\chi\mathrm{N}}(\alpha=1)=\frac{B}{48 \eta  \mu} P_{1}(A,\mu,\eta)
\label{eq:Cvd1}
\end{equation}
where $P_{1}(A,\mu,\eta)$ is given by
  \begin{align}
 &  = \sqrt{\pi } \bigg[1-2 \text{erfc}(\eta )+\text{erf}(A-\eta )+\text{erfc}(A+\eta )\bigg]
  \nonumber \\
 &  \times \bigg[\mu  \left(4 \left(A^2-3\right) \eta ^2+2 A^2-4 \eta ^4-3\right)  \nonumber \\
 &
 +2 A^2 \left(2 A^2-2 \eta ^2-1\right) \mu_{-}^2 \bigg]
 \nonumber \\
& + \eta \mu  \left(2 \eta ^2+5\right) \left(e^{4 A \eta }-2 e^{A (A+2 \eta )}+1\right)
\nonumber \\
 & +2 A^3 \left(e^{4 A \eta }-1\right) \mu_{-}^2
  \nonumber \\
& + 2 A^2 \eta  \bigg[\left(e^{4 A \eta }+1\right) \mu_{-}^2+2 e^{A (A+2 \eta )} \left(\mu -\mu_{-}^2\right)\bigg]
 \label{eq:fvd1}
 \end{align}
 
\subsection{For $\alpha=2$ (i.e., $\sigma\propto v^{4}_{\mathrm{rel}}$) case}
 For $\alpha=2$, an analytic expression of $C_{\chi\mathrm{N}}$ is given by 
\begin{equation}
C^{\mathrm{vd}}_{\chi\mathrm{N}}(\alpha=2)=\frac{B}{144 \eta  \mu ^2} P_{2}(A,\mu,\eta)
\label{eq:Cvd2}
\end{equation}
$\mathrm{where} \quad  P_{2}(A,\mu,\eta)  =$
  \begin{align}
& \sqrt{\pi }\bigg[(1-2 \text{erfc}(\eta )+\text{erf}(A-\eta )+\text{erfc}(A+\eta )\bigg]
  \nonumber \\
 &\times \bigg[\mu ^2 \left(A^2 \left(8 \left(\eta ^2+3\right) \eta ^2+6\right)-8 \eta ^6-60 \eta ^4-90 \eta ^2-15\right) 
 \nonumber \\
& +4 A^2 \mu_{-}^2  \bigg\lbrace\left(A^2 \left(4 \eta ^2+2\right)-4 \eta ^2 \left(\eta ^2+3\right)-3 \right) \bigg\rbrace \mu
 \nonumber \\
& + (-1+2A^{2}-2\eta^{2})\mu_{-}^2 \bigg]
 \nonumber \\
& +2e^{-(A+\eta )^2}\bigg[A \mu^2 \left(4 \eta^{2}\left(\eta ^2+6\right)+15\right) \left(e^{4 A \eta }-1\right) 
\nonumber \\
&+ \eta \mu^2 \left(4 \left(\eta ^2+7\right) \eta ^2+33\right) \left(e^{4 A \eta }-2 e^{A (A+2 \eta )}+1\right)
\nonumber \\
&+4 A^5 \left(e^{4 A \eta }-1\right) \mu_{-}^4+4 A^3 \mu  \left(e^{4 A \eta }-1\right) \left(\mu +\left(2 \eta ^2+3\right) \mu_{-}^2\right)
  \nonumber \\
&  +4 A^4 \eta  \mu_{-}^2 \left(\left(e^{4 A \eta }+1\right) \mu_{-}^2-2 e^{A (A+2 \eta )} \left(\mu_{-}^2-2 \mu \right)\right)
  \nonumber \\
&  +4 A^2 \mu \eta \left(e^{4 A \eta }+1\right) \left(2 \mu +\left(2 \eta ^2+5\right) \mu_{-}^2\right)  \left(2 \eta ^2+5\right) e^{A (A+2 \eta )} \nonumber\\ 
& 
\left(\mu -2 \mu_{-}^2\right) \bigg]
 \label{eq:fvd2}
 \end{align}
\subsection{For $\alpha=-1$ (i.e., $\sigma\propto v^{-2}_{\mathrm{rel}}$) case}
In this case, no analytic solution has been obtained. Therefore, we adopt numerical method to estimate the $C^{\mathrm{vd}}_{\chi\mathrm{N}}(\alpha=-1)$.

\section{Derivation of the accreted DM particles via MDCS}
\label{momdep}
In the MDCS case, the cross-section, $\sigma^{\mathrm{md}}_{\chi\mathrm{N}}\propto q^{2\beta}$. Here we assume $q\simeq \sqrt{2}\left(\frac{m_{\chi}m_{\mathrm{N}}}{m_{\chi}+m_{\mathrm{N}}}\right)v_{\mathrm{esc}}$ which suggests that the cross-section is independent of the initial DM velocity. Further, $\Omega^{-}(w)$ in this case is given by  
\begin{equation}
 \Omega^{-}(w)=n_{\mathrm{N}}\sigma^{\mathrm{md}}_{\chi\mathrm{N}}F(q^{2})~w~\frac{\mu^{2}_{+}}{\mu}\left( \frac{\mu}{\mu^{2}_{+}} - \frac{u^{2}}{w^{2}}\right) \Theta \left( \frac{\mu}{\mu^{2}_{+}} - \frac{u^{2}}{w^{2}}\right)~.
  \label{eq:omegawmd}
 \end{equation}
Now, rewriting the above equation in terms of the dimensionless variable, we obtain 
 \begin{equation}
 \Omega^{-}(w)= n_{N}\sigma_{\chi\mathrm{N}} \left(\frac{q}{q_{0}}\right)^{2\beta}~F(q^{2})\frac{1}{w}~\frac{v^{2}}{A^{2}}~\left(A^{2} -x^{2} \right)\Theta(A-x).
 \label{eq:omegawmd1}
 \end{equation}
Further, using Eq. (\ref{eq:Cnx}), Eq.(\ref{distx}), and Eq.  (\ref{eq:omegawmd1}), we calculate the captured rate of the DM particles.  
For arbitrary power of $\beta$, a analytic expression of $C^{\mathrm{md}}_{\chi\mathrm{N}}$ is given by
\begin{equation}
C^{\mathrm{md}}_{\chi\mathrm{N}}(\beta)=\left(\frac{q}{q_{0}}\right)^{2\beta}~C^{\mathrm{const}}_{\chi\mathrm{N}}~.
\label{eq:Cmd1}
\end{equation}
\bibliographystyle{utphys}
\bibliography{BDMinNS}
 
\end{document}